# The La Silla – QUEST Kuiper Belt Survey

David Rabinowitz[1], Megan E. Schwamb[1,2], Elena Hadjiyska[1], Suzanne Tourtellotte[3]

[1] Center for Astronomy and Astrophysics, Yale University, P.O. Box 208120, New Haven CT, 06520-8120 USA
[2] NSF Astronomy and Astrophysics Postdoctoral Fellow
[3] Dept of Astronomy, Yale University, P.O. Box 208101, New Haven CT, 06520-8101 USA

short title: The La Silla – QUEST KBO Survey

ABSTRACT

We describe the instrumentation and detection software and characterize the detection efficiency of an automated, all-sky, southern-hemisphere search for Kuiper Belt objects brighter than R mag 21.4. The search relies on Yale University's 160-Megapixel QUEST camera, previously used for successful surveys at Palomar that detected most of the distant dwarf planets, and now installed on the ESO 1.0-m Schmidt telescope at La Silla, Chile. Extensive upgrades were made to the telescope control system to support automation, and significant improvements were made to the camera. To date, 63 new KBOs have been discovered, including a new member of the Haumea collision family (2009 YE7) and a new distant object with inclination exceeding 70 deg (2010 WG9). In a survey covering ~7500 deg$^2$, we have thus far detected 77 KBOs and Centaurs, more than any other full-hemisphere search to date. Using a pattern of dithered pointings, we demonstrate a search efficiency exceeding 80%. We are currently on track to complete the southern-sky survey and detect any bright KBOs that have eluded detection from the north.

## 1. INTRODUCTION

The year 2008 marked the conclusion of two surveys for Kuiper Belt Objects (KBOs) conducted with the 160-Megapixel QUEST camera (Baltay et al. 2007) on the 1.2-m Oschin Schmidt telescope at Palomar (Brown 2008, Schwamb, Brown, & Rabinowitz 2009, Schwamb et al. 2010). These and an earlier Palomar survey (Trujillo & Brown 2003) spawned one of the most exciting periods in the exploration of the outer solar system. With a sky coverage of ~20,000 deg$^2$ searched to R-band magnitude limit $M_L$=20.5 and ~12,000 deg$^2$ searched a second time to $M_L$= 21.3, these surveys exposed a new population of dwarf planets with sizes close to that of Pluto (Brown, Trujillo, & Rabinowitz 2004, 2005, Rabinowitz et al. 2006, Brown 2008, Sicardy et al. 2011), ultimately leading to a new definition that excluded Pluto from the list of major planets (Weintraub 2007). Earlier surveys covering much smaller areas to fainter limits had already detected hundreds of fainter KBOs, and established that Pluto follows a typical Kuiper Belt orbit (Jewitt & Luu 2000, Millis et al. 2002, Larsen et al. 2007, Kevelaars et al. 2009). But the much larger size and brightness of Pluto relative to its dynamical cousins preserved its place among the planetary pantheon. Only with the discovery at Palomar of other bodies like Pluto massive enough to gravitationally retain a surface rich in volatile ices did it become clear that Pluto was not unique. But of more importance to our knowledge of the origin and evolution of the solar system are the physical properties observed for the dwarf planets. With their unusual surface compositions compared to smaller KBOs and their distinctive satellite systems, they provide us with a new understanding of KBO composition and structure, volatile retention, collisional dynamics, surface weathering, and dynamical evolution (Morbidelli & Levison 2004, Rabinowitz et al. 2006, Brown et al. 2006, Schaller & Brown 2007a, 2007b, Brown et al. 2007, Dumas et al. 2007, Sheppard 2007, Rabinowitz et al. 2008, Brown 2008, Levison et al. 2008, Abernathy et al. 2009, Leinhardt et al. 2010, Lykawka et al 2012).

Here we describe a new survey in the southern hemisphere that further explores the population of large KBOs. Our survey is intended to cover the entire sky south of the ecliptic (except the galactic plane) to depth $M_L \sim 21.5$, including areas not accessible to or not completely searched by previous northern-hemisphere searches. To date, no other KBO surveys of comparable scale and sensitivity have been conducted in the South. Although previous northern-hemisphere searches have covered declinations as far south as -25 deg, the coverage southward of ecliptic latitude -20 deg is mostly incomplete at our target depth. Recently, Sheppard et al. (2011) reported the results of a 2500-deg$^2$ search to limit $M_L = 21.6$ south of Dec -25 deg, partially covering the galactic plane. They found fourteen new KBOs with the largest several hundred km in diameter, but none rivaling Pluto in size. Bannister et al. (2011) reported a search of 9500 deg$^2$ to $M_L = 19.5$ yielding no dwarf planets. However, based on the discovery rate of large KBOs at Palomar and their wide-ranging latitudes (the Palomar surveys detected 8 KBOs larger than ~1000 km at latitudes ranging uniformly -18 to + 30 deg), we expect at least a few similar-sized KBOs remain undiscovered in the south.

Our new survey uses the same QUEST camera used for the Palomar KBO surveys and for the Palomar-Quest survey for supernova, transients, and quasars (Copin et al. 2006, Djorgovski et al. 2008, Bauer et al. 2009a, 2009b, 2009c, 2012). The camera has now been moved from the 1.2-m Oschin Schmidt to the 1.0-m Schmidt telescope of the European Southern Observatory (ESO) at La Silla, Chile. Half the dark time is dedicated to the KBO survey, with the remaining time dedicated to a search for nearby supernovae, variable stars, and other transients (Rabinowitz et al. 2011a, Hadjiyska et al. 2012, Zinn et al. 2012). Significant changes were made to improve the cooling system and a new wide band-pass filter was fabricated to eliminate low-level fringes in the images caused by night-sky lines at red wavelengths. A major upgrade of the ESO Schmidt was undertaken to enable fully automatic operation and high-speed transmission of image data to institutions within the USA. New software is used to process and search the survey images, to keep track of image quality, and to schedule nightly search fields. And a new search strategy is being implemented to improve the efficiency of the search.

In this paper we describe the details of the hardware upgrades and new search software and characterize the detection efficiency of the KBO search. With significantly darker skies, better seeing and improved tracking at La Silla compared to Palomar, and with a revised pointing strategy, we are on track to complete a full southern-hemisphere search for KBOs and other distant objects with magnitude limit $M_L = 21.4$ and efficiency exceeding 80%.

## 2. INSTRUMENTATION

### 2.1 *The ESO 1.0-m Schmidt Telescope*

With the termination of the Palomar KBO and transient surveys in 2008, it was natural to look for an appropriate telescope in the south to complete an all-sky survey for the largest KBOs and to continue the search for supernovae, variables, and other transients. The ideal telescope would have a Schmidt design, like the Oschin telescope, since the QUEST camera was specifically designed for this configuration. Such telescopes have a primary mirror of spherical shape, larger in size than the telescope's entrance aperture, providing an extraordinarily wide field of view. But the design requires a correcting lens of complex shape spanning the entire entrance aperture. Fabrication techniques limit the maximum size of these lenses to ~1m. Only a few Schmidt telescopes with apertures larger than 1 m have ever been constructed, including the 1.2-m telescopes at Palomar and the Australian Astronomical Observatory, and the 2-m telescope at the Karl Schwarzschild Observatory, Germany.

Fortunately for our program, the ESO 1.0-m Schmidt was available. The telescope is one of the largest Schmidt configurations in the southern hemisphere, situated at a dry site with dark skies and excellent seeing, and had been sitting idle since decommissioning in 1998. Having a nearly identical optical configuration to the Palomar Schmidt (but smaller entrance aperture), the QUEST camera could be installed without any changes to its front-end optics. ESO provided Yale with the opportunity to move the QUEST camera to the ESO Schmidt with the expectation that Yale would update the control system for automated operation, replace the plate holder at the prime focus with the 160-Megapixel QUEST camera, and establish an independent internet connection for transfer of the image data to Yale. Yale began to implement these changes in early 2009, and was ready for remote and automated operation of the ESO Schmidt by August of 2009.

The upgrade of the control system required a complete replacement of the controlling electronics for both telescope axes and the focus hub, including servo amplifiers, encoders, the controlling computer and control software. However, no changes were made to the existing drive motors, nor to the motors and relays controlling the rotation of the dome and operation of the dome shutter. Yale installed new limit switches to better control the operation and remote monitoring of the dome shutter. To control the telescope locally, operators use a menu-driven program running on a small computer in the control room of the Schmidt dome. In practice, however, all operations are automated, with commands transmitted externally through a digital interface to the controller.

For automated operation, we use the same system previously installed at the Palomar Schmidt telecope (Baltay et al. 2007). Each night, various programs run simultaneously on computers in the Schmidt control room to sequence telescope pointing and camera exposure, all coordinated by a master scheduling program. The scheduling program receives a list of requested pointings and exposures uploaded daily from Yale. As at Palomar, a remote operator of one of the larger telescopes at the site (the ESO 3.6m) decides when conditions are suitable for opening the Schmidt, and sends a remote command each night to enable the control software to open the dome. The control software automatically closes the dome whenever another nearby telescope (the 2.2-m) is closed, or when the sun rises, or when the remote operator sends a command to close. The remote operator can monitor and control the state of the Schmidt telescope via a web-based interface.

Because we must process each night of survey data at Yale the day after acquisition, we require a communication link to La Silla that can handle ~100 Gbytes per night. Unfortunately, no such link was available at La Silla. Yale therefore installed a private, high-speed radio link between La Silla and Cerro Tololo, where there is direct 100-km line of sight. Because no commercial radios are available that can handle transmissions over that distance, programmable radios were installed with experimental software developed for long-distance links in developing countries (Leffler 2009). The radio link achieves a bit rate of ~20 Mbits/sec, sufficient to transmit the survey data to Yale as fast as they are acquired. The receiving antenna at Cerro Tololo connects to a high-speed internet backbone that is linked to the US mainland (Kennedy 2002).

2.2 *The 160-Megapixel QUEST camera*

When the QUEST camera was installed at the ESO Schmidt, no changes were made to the focal plane elements or the readout electronics. Hence, the intrinsic characteristics are the same as described earlier for the Palomar - QUEST survey (Baltay et al. 2007). Briefly, the camera consists of 112 backside-illuminated and thinned charge-coupled devices (CCDs) originally designed for drift scanning. Each CCD is a 600 x 2400 pixel array, with pixel scale 0.88" and peak quantum efficiency of ~95%. Four separate structures (fingers A, B, C, and D) support the CCDs inside the camera dewar (see Fig. 1), with each finger capable of rotating to precisely align the parallel shift direction of the CCDs to the arcs traced by drifting stars in the focal plane. At La Silla, the camera is oriented to allow drift scanning, but we do not use this readout mode for the KBO survey. As with the survey at Palomar, we use a standard stare mode for all imaging (i.e. the telescope tracks the target field during the exposure, and the camera shutter is closed during readout). Readout time is ~40 seconds. Eleven of the CCDs do not function, and these are shown as blank areas on the fingers in Fig. 1. The remaining CCDs vary in quality. Several have regions of high dark currents owing to defective, light-emitting pixels in the CCD substructure or in a corner readout amplifier. By subtracting dark calibration images, the adverse influence of these defects on source detection is largely eliminated. The sky area covered by the functioning CCDs is 8.65 $deg^2$.

A significant improvement in the camera image quality was obtained when replacements were made to the power supply for the readout electronics and to the optical-fiber interface between the control computer and the camera. These changes eliminated digital noise that affected the Palomar data. The noise occurred whenever the pixel values in any of the CCD images were near particular large-valued, integer powers of 2. Near these values, the pixel values did not increase in proportion to the source brightness until the source intensity increased by a large percentage. Although photometric measurments were possible, the resulting noise limited the precision. With this noise now removed, higher photometric precision can now be achieved at La Silla than was possible at Palomar.

The most significant improvement to the camera in the move to La Silla was a change to the method of cooling the focal plane. The previous system required large amounts of liquid nitrogen continuously supplied to a dewar situated behind the camera inside the telescope body. At La Silla, this dewar is now replaced with a pair of 60-Watt cryo-refrigerators. Each unit transfers heat using a Gifford-McMahon cycle, where compressed helium at room temperature drives the motion of a piston, thereby drawing thermal energy from the cold head at the end of the piston cylinder. Two large helium compressors sit on the dome floor beneath the telescope, connected by long, flexible high-pressure lines to the piston heads. Recirculated water cools each compressor, with the heat dissipated outside the dome by air/water heat exchangers. To minimize vibration of the focal plane induced by the piston cycling, the two cryo-refrigerators are firmly bolted to the stiff spider vanes supporting the focus hub of the telescope. Because the units mechanically connect to the camera head only by flexible copper straps and flexible vacuum housings, their vibrational energy is almost completely absorbed by the large mass of the telescope. There is no detectable influence of the vibration on image quality.

2.3 *The Qst Survey Filter*

At Palomar, a significant amount of image noise was caused by interference patterns in the CCDs associated with night-sky emission lines at red wavelengths. At La Silla, we have largely eliminated these patterns using an optical interference filter with nearly uniform transmission from 400 to 700 nm (see Fig. 2). We call this the "$Q_{st}$" filter. It covers the range of wavelengths where the camera is most sensitive to the reflected sunlight from solar system bodies, yet minimizes the contribution of scattered moonlight which peaks at bluer wavelengths. Because we were unable to obtain a single filter large enough to cover the whole focal plane (~30 cm diameter), we use four square filters butted together at their edges. A narrow opaque mask covers the edges where the filters meet. This prevents light from scattering off the filter edges. However, there is some vignetting in the focal plane behind and near the masked area. This occurs over a span of ~300 pixels at the abutting sides of the two central CCDs in each column, where the average light throughput is reduced by ~75% .

3. SEARCH METHOD

3.1 *The Search Strategy*

We use a search strategy at La Silla based on that of the Palomar searches. To attain the desired depth and efficiency, we expose each field for 180 s with three passes. All fields must be observed at air mass < 2 with a 2-hour interval between passes in most cases, but we accept fields imaged at least 1.5 h apart (owing to observing interruptions or scheduling conflicts). We target all fields only when they are within 30 deg of the sun's opposition calculated for the new moon date of the lunation. This geometry maximizes the apparent retrograde motion owing to parallax from the Earth's relative motion (~1"/hr for objects at 150 AU). We also fix all field locations to a grid covering the sky, and cover all positions with a pair of pointings dithered by 0.5 deg in RA (we refer to these as dither pair "A"). This is the same grid and dither pattern that was used at Palomar. With the grid points separated by 4 deg in RA and 4.5 deg in Dec, this ensures complete coverage of the sky except for gaps between the CCDs on each finger, and for gaps caused by non-working CCDs. To fill in these gaps, we have augmented the Palomar strategy by adding a second dithered pair to each grid position (dither pair "B"), displaced from the first pair by 0.416 deg in Dec (2.5 times the Dec spacing of the CCDs). Figure 1 shows the direction and magnitude of the offsets required for the A and B pairs. Schwamb et al. (2010) used a similar dither pattern for some of their survey fields.

We schedule fields nightly, with priorities based on ecliptic latitude. Highest priority goes to those latitudes with the highest density of bright KBOs, but also where new discoveries are most likely. In order of increasing priority, we therefore search the sky in four latitude bands: -10 to -40, -40 to -60, -10 to 0, and -60 to -80 deg. Highest priority goes to latitude range -10 to -40 deg for three reasons: this range has not been well targeted in previous searches; the number of KBOs brighter than our detection limit does not diminish significantly at these latitudes compared to the ecliptic (Brown 2008, Schwamb et al. 2010); and most of the known dwarf planets were discovered more than 10 deg off the ecliptic (Eris, Makemake, Haumea, Sedna, and Orcus) . The range -40 to -60 deg has a much lower spatial density of bright KBOs, but we give it second priority because the area is relatively unexplored. We give third priority to range 0 to -10 deg because the region has already been extensively explored, but not completely. We give the lowest priority to range -60 to -80 deg where expect the sky density of KBOs to be lowest. We strive to cover all the fields in all latitude bands with both dither pairs (A and B, see above discussion). However, we first target the fields in a given latitude band with only pair A, which covers most of the field area. Upon completion of the band, we go back and repeat it with pair B. We then go on to the next latitude band in order of priority. Once we cover all four bands with both dithers pairs, and if observing time is available, we cover the latitude range -80 to -90 deg and regions north of the ecliptic up to +20 deg Dec.

We conduct the survey in a wide variety of sky conditions, not restricted to photometric nights. To ensure that we reach our target survey depth, we assess the limiting magnitude of every search field as follows. For each exposure, we independently measure a magnitude zero point for five good-quality CCDs in separate areas of the focal plane with reference to the R-band magnitudes of field stars listed in the USNO A2.0 catalog (Monet et al. 1998, Assafin et al. 2001). The zero point uncertainty is limited by the catalog uncertainty of 0.3 for stars with magnitudes greater than 17. Unfortunately, there is no more precise photometric catalog covering the majority of our survey area. The overlap with the Sloan Digital Sky Survey is only ~6% (Adelman-McCarthy et al. 2008). For each CCD, we then combine R-band magnitudes measured for all the stars from all three survey passes and determine the "peak magnitude", which is the magnitude with peak frequency (measured in bins with resolution 0.2 mag). If the median peak magnitude of the five CCDs is greater than or equal to 20.6 mag (corresponding roughly to a detection limit of $M_R$ =21.5, see below), we accept the field. We reschedule rejected fields for new observations on a subsequent night. Note that these estimates of our detection limit are dependent on the luminosity function of the stars, which varies with galactic latitude. However, given the expected variations (Bahcall & Soneira 1980), we calculate a shift of no more than +/- 0.14 mag in our measured peak magnitudes. Hence, the precision of the USNO catalog is the limiting factor in our measurement.

3.2 *The Detection Pipeline*

The detection pipeline consists of a series of programs that run automatically on computers at Yale each morning to process the previous night's images uploaded from La Silla. As with most solar-system survey programs (for example, Rabinowitz 1991, Trujillo & Jewitt 1998, Pravdo et al. 1999, Millis et al. 2002, Trujillo & Brown 2003, and Petit et al. 2004), the pipeline identifies moving objects by cataloging source positions in each of three passes and then comparing the catalogs to isolate the sources changing position at a constant rate. All images are dark subtracted and flat fielded using average darks and twilight flats collected each evening. SExtractor (Bertin & Arnouts 1996) is used to isolate and measure source coordinates, with the requirement that pixel intensity of each source exceed 1.5 times the background noise in the image. Candidate moving objects are then identified by their positions relative to field stars in the three catalogs. Astrometric and photometric solutions are then determined with reference to field stars listed in the USNO-A2.0 catalog. The pipeline uses routines from astrometry.net (Lang et al. 2010) for initial astrometric solutions, but refines these solutions as necessary to minimize the resulting astrometric residuals. Photometric zero points are determined separately for each CCD image in the system of the USNO R band.

To reduce the number of false detections, we impose the following restrictions on the candidate detections:

1. The uncertainty of the measured brightness must not exceed 0.5 mag (equivalent to a minimum signal-to-noise ratio of 2.9 times the sky noise).

2. The displacement between consecutive passes must exceed 1.5 pixels (i.e. exceed 0.0044 deg/day for fields observed with 2 h separation). This rejects sources more distant than ~200 AU.

3. The motion must not exceed 0.10 deg/day in ecliptic longitude or latitude. This rejects valid sources at opposition closer than about 15 AU.

4. An orbital fit to the measured motion must yield residuals with $\chi^2 < 10$. Here we use the orbit-fitting programs supplied by Bernstein & Khushalani (2000) . This $\chi^2$ limit is well above the range of values we obtain for valid detections (usually less than 2).

5. The fitted distance to the candidate must exceed 10 AU.

6. The measured brightness values must agree to 1 magnitude.

With the above cuts, the number of false detections per night is usually < 100, depending on the variation in seeing, extinction, and the number of stars per field. Hence, we must visually inspect images of each candidate to decide validity. For this purpose, the pipeline creates small, animated snapshots showing the motion of each source relative to the field stars. The pipeline also creates a static 3x3 image mosaics showing the three positions of each candidate at the three different observation times (see Pravdo et al. 1999). Viewed with a web browser, the validity of most candidates is obvious within seconds of viewing.

For example, figure 3 shows the discovery images of 2010 WG9 (top row, first three images left to right). These would normally be combined into a single animated snapshot, clearly revealing the motion for most objects. The last image in the row is the 3x3 mosaic, assembled from 3 sub-images of each discovery image. A valid source should appear centrally in the diagonal subimages (top left to bottom right), corresponding to its positions in the three discovery images (p1, p2, and p3) at the respective times of these observations (t1, t2, and t3). Also, no source should appear in the remaining subimages, corresponding to position p1 at times t2 and t3, p2 at times t1 and t3, and p3 at times t1 and t2 (left to right, top to bottom). Clearly, 2010 WG9 satisfies these criteria.

The bottom row of Fig. 3 shows a typical false detection. There is no source with clear motion in the discovery images. Very faint sources appear centrally along the diagonal of the mosaic, but also in some of the off-diagonal positions. It is likely the pipeline has marginally detected a very faint star in one of the discovery images and failed to detect it in the others. It has also detected some noise in the other images coincidentally appearing at positions consistent with the displacement of a source moving at a constant rate.

3.3 *Follow-up Observations*

We follow up all of our discoveries with the 1.3-m telescope of the Small and Medium Aperture Research Telescope System (SMARTS) at Cerro Tololo or with the ESO Schmidt telescope, measuring positions over a large enough time span (usually a few months) to secure the orbits. All our discovery and follow-up positions are reported to the Minor Planet Center (MPC) within a few weeks of acquisition. We also measure colors and monitor variability for some of the brighter targets. When possible, we obtain pre-discovery images using archived images from our own survey, the La Silla – QUEST transient survey, or the Palomar surveys.

## 4. SEARCH CHARACTERISATION

### 4.1 *System Performance*

After a breaking-in period of ~6 months, the telescope, camera, control software, and communication system have been stable and performing well. Median seeing has been 2.0" for all 180-sec exposures meeting our depth requirement (see below). Unguided tracking stability is ~2" in 5 min, and with night-to-night pointing precision of ~10". The only significant problem we have had is with the dome rotation. Owing to the aging of load-bearing rubber elements in the original dome support wheels (and spares), the telescope has been out of operation for significant intervals while ESO engineers have made temporary repairs to the wheels. Of the 300 nights available to the KBO survey since 2010 Sept, only 100 could be used owing to bad weather and dome problems. Recently, ESO completely replaced the original wheels with a system of bogies designed and fabricated in house. The new system is now working without problems.

### 4.2 *Magnitude Limit and Efficiency*

For the purpose of population estimates and to gauge the effectiveness of our survey, we need to determine the R-band magnitude limit, $M_L$, and measure two different efficiencies, $\varepsilon$ and $\kappa$, as a function of the observed R-band magnitudes, $M_R$. Here, we define $\varepsilon$ to be the single-pointing efficiency, which is the likelihood of detecting a KBO in a given field if the KBO appears within the sky footprint of the functional CCDs (see Fig. 1). Any effort to use our survey results to constrain the KBO population requires a measure of $\varepsilon$ to account for the selection bias of the survey. We define $\kappa$ to be dithered-pointing efficiency, which is the likelihood of detecting a KBO at least once if it is located anywhere within the overlap area of a complete set of dithered fields (pairs A and B, see above). Here we do not require the KBO to fall within the area bounded by any particular CCD, only that it falls within the outer perimeter of the overlapping fields. We expect $\kappa > \varepsilon$ owing to the multiple chances for detecting a KBO with 4 dithered fields, and because the areas covered by the least sensitive CCDs in one pointing will be covered by adjacent CCDs of higher sensitivity when the pointing is dithered. We define limit $M_L$ to be the threshold value for $M_R$ at which $\varepsilon$ drops to less than half its average value at bright magnitudes.

We measure $\varepsilon$ and $M_L$ using two methods. Both rely on counting the fraction of known objects that appear in our fields and that our pipeline detects. In the first method, we target main-belt asteroids specifically because they are numerous at bright magnitudes, where they provide a good measure of $\varepsilon$. A complete set of 4 dithered pointings near opposition yields a sufficient number for this purpose (~650). At faint magnitudes, however, there are too few main-belt asteroids with well-determined orbits to accurately characterize $M_L$. Furthermore, their motion during a 180-s exposure is comparable to our typical seeing, leading to significant trailing loss. In the second method, we target known KBOs, which are relatively numerous at faint magnitudes. Counting all the known KBOs that appear in all our survey fields, there are a sufficient number to measure $M_L$, but not enough to characterize ε at bright magnitudes. Combining the observations from both methods, however, we are able to measure the magnitude-dependence of ε over the widest possible range. We can also use the observations of main-belt asteroids to measure κ.

To target main-belt asteroids, we observed a particular field near opposition on a clear night with seeing ~2.5", slightly worst then the median seeing of our survey fields (2.2"). The limiting magnitude, determine from the peak of the stellar magnitude distribution (see Sec 3.1) was 21.2, nearly the same as the median limit for our accepted survey fields (21.0). We note, however, that the resulting measurement for ε at bright magnitudes does not depend significantly on the quality and depth of these fields. As will be shown below, ε is magnitude independent for $M_R < M_L - 0.5$. Even if these fields were extincted by 0.5 mag, we would measure the same bright-magnitude efficiency.

We recorded a complete, 3-pass sequence with dither pairs A and B and processed the data with our normal detection pipeline. Exposure times were the same as for the KBO search. The only changes from our normal survey strategy were (1) to shorten the interval between passes from 2.0 to 0.25 h, and (2) to admit asteroids with motions up to 1 deg/day with no minimum distance threshold. With the shorter time interval, the typical displacement for the main-belt asteroids at opposition ranges from 4" to 12". This overlaps the range for typical KBOs in our normal cadence. For example, a typical KBO at 50 AU moves ~6" in 2 h, and the most distant KBO (Eris at 100 AU) moves 3". Hence, the external factors affecting our efficiency (source displacement, detector noise and sensitivity, variations in point-spread function with field position, star crowding, and detection algorithm) were the same for the observations of the main-belt asteroids in this calibration sequence as they are for the KBOs in our normal survey.

The only significant factors biasing the result are the slight trailing of the main-belt asteroids owing to their faster motion relative to KBOs and the poorer seeing of the calibration fields. In a 180 sec exposure, a main-belt asteroid moves from 0.9" and 2.4", an amount comparable to the seeing, whereas the motion of a KBO is insignificant. The resulting trailing loss is 0.23 mag, owing to a decrease in the peak-pixel flux in the source image. We measured this decrease by digitally shifting the calibration images and then coadding to yield a summed image whose field stars are trailed by the expected amounts for the asteroids. We also measure a 0.22 mag loss relative to our average survey fields owing to the relatively poorer seeing. This we determined by measuring the variation in peak-pixel flux as a function of the variable seeing in the different calibration images. Together, the two losses should lead us to underestimate $M_L$ by ~0.45 mag. As discussed above, these effects do not alter our bright-magnitude measurement for e. However, they justify ignoring the result for $M_R > M_L - 0.5$.

For a given magnitude range, we then measure ε for each of the 4 pointings individually by evaluating $N_{obs}/N_{pred}$, where $N_{pred}$ is the number of asteroids we predict within each of the sky areas bounded by the functional CCDs and $N_{obs}$ is the number of matching asteroids we detect. Here, we restrict the count to only those asteroids cataloged by the MPC with numerical designations. Because we can predict their positions to a precision of a few arcsec, this restriction eliminates any ambiguity in the matches. For each such asteroid within a 5 deg radius centered on the 4 pointings, we then use JPL Horizons[1] to compute the position and apparent magnitude, $M_{pred}$, at the precise epoch that we observed the second pass. To measure κ, we combine the detections and predictions for all 4 pointings and evaluate $N_{obs}/N_{pred}$, where $N_{pred}$ is now the number of distinct asteroids we predict anywhere within the perimeter enclosing the sky area overlapped by the 4 pointings, and $N_{obs}$ is the number of matching asteroids we detect at least once. Here we count multiple detections of the same asteroid as a single detection.

To measure ε using KBO observations, we repeated the above calibration substituting all of our actual survey fields for the main-belt calibration fields and substituting all the previously known KBOs with well-determined orbits (objects with at least a 1 year observational arc) for the known main-belt asteroids. We extract the targets from the listing by the MPC of known distant objects, excluding our own discoveries. To determine $N_{obs}/N_{pred,}$ we count only those KBOs with predicted positions inside the areas of sky covered by the functional CCDs.

Figure 4 shows the resulting values for ε versus $M_R$ determined from main-belt asteroids (open squares) and from KBOs (filled squares). The error bars show the Poisson counting error, $N_{obs}^{1/2}/N_{pred}$., where we have counted the detections in 0.5-mag bins. Note that we can not measure $M_R$ for all the predicted asteroids in the field, just for those we detect. The only available brightness measure for all the asteroids is $M_{pred}$. Comparing predicted and measured magnitudes for all the asteroids we do detect, we find an average difference $\Delta M_{corr} = <M_{pred} - M_R> = 0.163$ +/- 0.013 with rms variance 0.33. This variance is expected because $M_{pred}$ is determined from the absolute magnitudes cataloged by the MPC. These are sometimes based upon reported magnitudes that are not well calibrated. Also, each asteroid will have unpredictable brightness variations due to rotational modulation and opposition surges. To have a consistent measure of $M_R$ for all the predicted asteroids, we therefore take $M_R = M_{pred} - \Delta M_{corr}$. For main-belt asteroids, the resulting values for $N_{pred}$ exceed 10 for all magnitude bins except $M_R > 21.5$, where none are predicted, and $M_R < 16.5$, where $N_{pred}$ ranges from 4 to 9. For KBOs, $N_{pred} > 10$ for all bins except $M_R = 19.5$, where only one is predicted, and $M_R < 19.5$, where none are predicted.

We fit the efficiency with a function, $\varepsilon(M_R)$, of the form

$$\varepsilon(M_R) = (\varepsilon_B/2)( 1 - \tanh[(M_R-M_L)/g] ) \qquad (1)$$

such that $\varepsilon = \varepsilon_B$ at bright magnitudes, $\varepsilon = \varepsilon_B/2$ at $M_R = M_L$, and ε falls to near zero over the range $M_L-g$ to $M_L+g$. Fitting the measurements for main-belt asteroids and KBOs separately, we find the values for $\varepsilon_B$, $M_L$, and g listed in table 1. These are the values that minimize $\chi^2$, with the uncertainty for each parameter determined by the range of values increasing $\chi^2$ by less than 1. We see that both main-belts asteroids and KBOs yield consistent values for $\varepsilon_B$ (0.76 +/- 0.04 and 0.69 +/- 0.11, respectively), but the value for $M_L$ from KBOs (21.48 +/- 0.11) is fainter than from main-belt asteroids (21.0 +/- 0.15) by

---

[1]   http://ssd.jpl.nasa.gov/?horizons

~0.5 +/- 0.2. This is consistent with the expected trailing loss for main-belt asteroids and the relative poor seeing of the calibration images (see discussion above). To obtain a fit that best characterizes the efficiency for KBO detection, we combine the results from main-belt observations at bright magnitudes, $M_R$ < 20.25, with the results from relatively faint KBO observations, $M_R$ > 20.25. This yields the curve plotted in figure 4 and the values for the data set labeled "combined" in table 1.

Figure 5 shows $\kappa$ vs $M_R$. Fitting the data with the same function we use to fit $\varepsilon$, we obtain the curve plotted in the figure, for which $\kappa$ has the limiting value 0.84 +/- 0.05 for bright magnitudes. As expected, this value is higher than the bright-magnitude efficiency, $\varepsilon_B$ = 0.76 +/- 04, we obtain for single pointings. At fainter magnitudes, the fit to $\kappa$ vs $M_R$ does not differ significantly from the single-pointing fit, but the uncertainty here is large owing to the small number of detections.

4.3 *Sky Coverage and Detection Rates*

From 2009 Dec to 2010 Jul we conducted a trial survey to test the equipment and software and to optimize our observing strategy and detection method. In this time we detected 40 distant objects (KBOs and Centaurs), all with distance R > 8 AU, of which 23 are new discoveries. Beginning 2010 Aug and up to the present (2012 May), we are conducting a more systematic survey. During this time we have acquired ~2000 exposures, each with 3 passes separated by at least 1.5 h, and each reaching our target depth, $M_L$ ~ 21.5. These observations cover 527 different fields, of which 421 are fully dithered (pairs A and B), and 90 are at least dithered with pair A. Figure 6 maps the field coverage, with completely dithered areas colored cyan, and all other areas colored blue. With each completely dithered field covering 17.85 $deg^2$, we have thus surveyed ~7500 $deg^2$ to our target depth with efficiency $\kappa$ = 0.84. The partial dithers cover an additional area of ~1600 $deg^2$, but with a lower efficiency (< 70%) owing to gaps in the array. We also observed many fields that do not meet our target depth or do not have the required time separation between passes.

Table 2 lists the following discovery information for all the distant bodies we have detected (105 as of 2012 May): the MPC designation, the first LSQ detection date (UT), the $M_R$ value we measure on the first detection date, the semi-major axis (a), eccentricity (e), inclination (i), absolute V-band magnitude (H), the observational arc and the orbit type. For nearly all objects, the orbital elements and H value are as listed by the MPC. However, for a few cases the observational arcs are very short (only a few days) and the MPC does not list the orbits. For these (indicated by an asterisk after the designation), we compute a, e, i, and H values from our observations using the fitting routines supplied by Bernstein & Khushalani (2000). The corresponding orbit type is the classification used by the MPC ("tno" for stable trans-Neptunian orbits and "scattered disk/Centaur" for everything else), except that we separated "scattered-disk/Centaur" into "scattered" and "centaur" based on the aphelion value (greater or less than 30 AU, respectively). We do not assign types to the short-arc orbits owing to their large uncertainty.

Of the objects listed in Table 2, we have detected 62 with our systematic sky coverage, and an additional 14 in the fields that did not meet our acceptance criteria (see table notes). Together, these comprise 42 new discoveries and 34 re-detections of previously known objects (including some of our own discoveries from the trial survey). These include several incidental detections of Sedna (2003

VB12) and Eris (2003 UB313), proving our capability for detecting objects at ~100 AU distance. We have followed nearly half of our discoveries for at least one opposition, and nearly all for at least 2 lunations.

Figure 7 shows the total number of detections (solid line) and our fractional sky coverage (dashed line) versus ecliptic latitude, counting only fields meeting our acceptance criteria, and only KBOs detected in those fields. Despite a fractional sky coverage ~50% or less at most latitudes, our total number of detections is already larger than reported for both of the recent Palomar surveys (52 by Schwamb et al. 2010 and 71 by Brown 2008). To properly compare the surveys requires a careful accounting of the limiting magnitude and sky coverage versus ecliptic latitude. We reserve this analysis for a later paper when our sky coverage is complete. Very roughly, however, from Schwamb et al. (2010) we expect a survey with 100% coverage to find ~100 KBOs with $M_R > 21$ at latitudes between 0 and -30 deg. With ~50% sky coverage in this range and with a slightly fainter magnitude limit, we have exceeded this expectation. This confirms that our search is progressing well, with an efficiency and magnitude limit comparable to what we have determined.

5. SUMMARY

We have described the instrumentation, search strategy, and detection method of a new, full-hemisphere, southern sky survey for KBOs. From target observations of main-belt asteroids, and from incidental detection of known KBOs, we demonstrate a deeper R-band magnitude and higher efficiency than any previous full-hemisphere survey. The survey is ongoing, and to date has covered only about half of the accessible sky south of the ecliptic (more than 15 deg from the galactic plane). Nonetheless, the number of KBOs we have detected is already larger than any previous full-hemisphere search. Of these detections, 63 are new KBOs that we are following to determine orbits and physical properties. One of these (2009 YE7) is a confirmed new member of the Haumea collision family (Trujillo et al. 2011). Two have unusually similar color and orbit (2010 FD49 and 2010 FE49), and may be members of a separated binary (Rabinowitz et al. 2011b). Another (2010 WG9) has an extraordinary inclination of 70 deg, indicating a possible origin as a returning Oort-cloud object (Schwamb et al. 2011). We plan to use the results of our survey to constrain models of Oort-cloud population and evolution (Brasser et al. 2012). Continuing the search to completion, we anticipate many more interesting discoveries, and look forward to the detection of the brightest remaining KBOs.

This research was supported by NASA grant NNX10AB31G and DOE grant DE-FG02-ER92-40704. MES is supported by an NSF Astronomy and Astrophysics Postdoctoral Fellowship under award AST-1003258. We also credit the following: Yale U. for supporting telescope operations; **Indiana U. colleague M. Gebhard and** Yale colleagues N. Ellman, W. Emmet, T. Hurteau, and R. Lauer without whose engineering, technical, and computer support we could not operate; ESO staff at La Silla G. Ihle, P. Francois, J. Fluxa, B. Ahumeda, A. Wright, and all the TIOS for outstanding instrument support; COMSOFT engineers D. Harvey and G. Stafford for their excellent upgrade the Schmidt control system; CTIO staff E. Figueroa, C. Smith, J. Hughes, and R. Lambert, USCD colleagues H-W. Braun and J. Hale, K. Fall at UC Berkeley, and S. Leffler for making the radio link possible; students E. Tuzinkiewicz, P. Fuentes-Huaiquilao and R. Ramirez for helpful efficiency studies; and finally Prof. C. Baltay at Yale for making the whole show possible.

Table 1.

Parameters for the best-fit efficiency function, $\varepsilon(M_R)$

| Data Set | $\varepsilon_B$ | $\varepsilon_B$ uncert. | $M_L$ | $M_L$ uncert. | g | g uncert. |
|---|---|---|---|---|---|---|
| main belt | 0.76 | 0.04 | 21.00 | 0.15 | 0.94 | 0.24 |
| KBO | 0.69 | 0.11 | 21.48 | 0.11 | 0.35 | 0.18 |
| combined | 0.74 | 0.04 | 21.44 | 0.11 | 0.39 | 0.19 |

Table 2.

Distant Objects Detected by the La Silla – QUEST Kuiper Belt Survey

2009 Dec to 2012 May

| Designation[a] | | Det. Date (yy mm dd) | | | $M_R$ | a (AU) | e | i (deg) | H | arc[b] | orbit[c] |
|---|---|---|---|---|---|---|---|---|---|---|---|
| **2009** | **YD7** | 9 | 12 | 16 | 21.5 | 129 | 0.9 | 30.8 | 9.9 | 3 | scattered |
| **2009** | **YE7**[d] | 9 | 12 | 17 | 21.1 | 44.58 | 0.14 | 29.1 | 4.4 | 3 | tno |
| 2007 | UK126[d] | 9 | 12 | 19 | 19.5 | 74.38 | 0.49 | 23.3 | 3.4 | 9 | scattered |
| **2009** | **YF7** | 9 | 12 | 19 | 21.7 | 12.06 | 0.46 | 31 | 10.7 | 3 | centaur |
| 2005 | RM43[d] | 9 | 12 | 21 | 19 | 92.23 | 0.62 | 28.7 | 4.4 | 11 | scattered |
| 1996 | TL66[d] | 9 | 12 | 23 | 20.2 | 84.84 | 0.59 | 23.9 | 5.4 | 10 | scattered |
| 2003 | UT292[d] | 9 | 12 | 23 | 21.4 | 39.59 | 0.3 | 17.5 | 6.9 | 6 | tno |
| **2009** | **YG19** | 9 | 12 | 25 | 20.9 | 62.62 | 0.52 | 5.2 | 5.9 | 47d | scattered |
| **2010** | **BL4** | 10 | 1 | 18 | 21.1 | 18.64 | 0.54 | 20.8 | 12 | 3 | centaur |
| **2010** | **EN65** | 10 | 3 | 7 | 20.4 | 30.72 | 0.32 | 19.2 | 6.9 | 11 | scattered |
| 2001 | FZ173[d] | 10 | 3 | 9 | 20.5 | 84.76 | 0.62 | 12.7 | 6.1 | 6 | scattered |
| **2010** | **EO65** | 10 | 3 | 9 | 21.1 | 24.97 | 0.44 | 11.4 | 9.5 | 3 | scattered |
| **2010** | **EP65** | 10 | 3 | 9 | 20.3 | 47.5 | 0.3 | 18.9 | 5.5 | 8 | tno |
| 2003 | FY128 | 10 | 3 | 10 | 20.2 | 49.32 | 0.25 | 11.8 | 4.9 | 10 | tno |
| **2010** | **EQ65** | 10 | 3 | 10 | 21.6 | 81.29 | 0.58 | 23.6 | 6.4 | 41d | scattered |
| **2010** | **ER65**[d] | 10 | 3 | 10 | 21 | 97.94 | 0.59 | 21.3 | 5.4 | 3 | scattered |
| 2004 | DW | 10 | 3 | 12 | 18.7 | 39.32 | 0.22 | 20.5 | 2.2 | 18 | tno |
| **2010** | **ES65** | 10 | 3 | 12 | 21.3 | 21.48 | 0.56 | 10.4 | 11.8 | 84d | scattered |
| **2010** | **ET65** | 10 | 3 | 13 | 20.9 | 62.2 | 0.36 | 30.6 | 5.2 | 6 | scattered |
| **2010** | **EU65** | 10 | 3 | 13 | 21.4 | 19.2 | 0.05 | 14.8 | 9.1 | 85d | centaur |
| **2010** | **FB49** | 10 | 3 | 17 | 21.2 | 22.5 | 0.2 | 24.4 | 7.5 | 3 | centaur |
| **2010** | **FC49** | 10 | 3 | 17 | 21.3 | 38.94 | 0.05 | 39.8 | 5.9 | 3 | tno |
| 1999 | DE9[d] | 10 | 3 | 19 | 20.1 | 55.49 | 0.42 | 7.6 | 5.1 | 10 | scattered |
| **2010** | **FD49** | 10 | 3 | 19 | 21.4 | 55.23 | 0.42 | 10.7 | 6.3 | 3 | scattered |
| **2010** | **FE49** | 10 | 3 | 21 | 21.4 | 53.59 | 0.37 | 11.7 | 6.5 | 3 | scattered |

| | | | | | | | | | | |
|---|---|---|---|---|---|---|---|---|---|---|
| 2010 | EK139[d] | 10 | 4 | 6 | 19.6 | 68.91 | 0.53 | 29.5 | 3.8 | 5 | scattered |
| **2010** | **GX34** | 10 | 4 | 9 | 20.7 | 28.61 | 0.42 | 11.6 | 8.2 | 3 | scattered |
| 2010 | EL139[d] | 10 | 4 | 14 | 20.6 | 39.26 | 0.07 | 23 | 5.1 | 2 | tno |
| **2010** | **GF65** | 10 | 4 | 14 | 21 | 36.18 | 0.42 | 12.3 | 7.2 | 62d | scattered |
| 2000 | GN171[d] | 10 | 5 | 12 | 20.1 | 39.22 | 0.28 | 10.8 | 6 | 9 | tno |
| 2002 | KX14 | 10 | 5 | 12 | 20.2 | 38.66 | 0.05 | 0.4 | 4.4 | 8 | tno |
| 2007 | JK43 | 10 | 5 | 12 | 20.2 | 46.05 | 0.49 | 44.9 | 7 | 4 | scattered |
| 2010 | HE79 | 10 | 5 | 12 | 20.6 | 38.84 | 0.18 | 15.8 | 5.2 | 5 | tno |
| **2010** | **JB80** | 10 | 5 | 12 | 21.5 | 12.91 | 0.2 | 17.8 | 11.7 | 57d | centaur |
| **2010** | **JC80** | 10 | 5 | 12 | 20.9 | 55.68 | 0.44 | 3.1 | 5.8 | 2 | scattered |
| 2009 | HH36 | 10 | 5 | 18 | 18.3 | 12.7 | 0.45 | 23.3 | 10.7 | 4 | centaur |
| 2003 | CO1 | 10 | 6 | 8 | 18.3 | 20.67 | 0.47 | 19.8 | 8.9 | 9 | scattered |
| 2007 | JJ43 | 10 | 6 | 8 | 20.7 | 47.78 | 0.16 | 12.1 | 3.8 | 6 | tno |
| **2010** | **LO33** | 10 | 6 | 8 | 20.7 | 24.03 | 0.37 | 17.8 | 8.1 | 100d | scattered |
| **2010** | **LN68** | 10 | 6 | 13 | 19.8 | 12.59 | 0.02 | 15.1 | 9.2 | 1d | centaur |
| 2004 | PF115 | 10 | 8 | 14 | 21 | 38.99 | 0.07 | 13.4 | 4.4 | 8 | tno |
| **2010** | **PK66**[e] | 10 | 8 | 14 | 21.4 | 40.93 | 0.01 | 13.6 | 5.6 | 80d | tno |
| **2010** | **PL66** | 10 | 8 | 14 | 22 | 20.92 | 0.37 | 24.4 | 8.4 | 55d | centaur |
| **2010** | **PT66** | 10 | 8 | 15 | 21.8 | 62.98 | 0.42 | 18.3 | 5.8 | 57d | scattered |
| 2007 | NC7[e] | 10 | 8 | 17 | 21.2 | 34.08 | 0.5 | 6.3 | 7.7 | 4 | scattered |
| **2010** | **RM45** | 10 | 9 | 5 | 22.4 | 60.95 | 0.57 | 18.9 | 6.5 | 94d | scattered |
| **2010** | **RF43** | 10 | 9 | 6 | 21.1 | 50.29 | 0.28 | 30.6 | 4.2 | 2 | scattered |
| **2010** | **RG43**[e] | 10 | 9 | 6 | 20.7 | 14.52 | 0.08 | 36.6 | 9.7 | 1d | centaur |
| **2010** | **RN45** | 10 | 9 | 7 | 21.7 | 39.95 | 0.1 | 32.9 | 5.9 | 66d | tno |
| **2010** | **RE64** | 10 | 9 | 9 | 21.3 | 63.71 | 0.42 | 13.5 | 4.3 | 2 | scattered |
| **2010** | **RF64** | 10 | 9 | 9 | 21 | 42.37 | 0.15 | 28.7 | 5.7 | 29d | tno |
| **2010** | **RM64**[e] | 10 | 9 | 9 | 19.5 | 19.79 | 0.69 | 27 | 11.1 | 2 | scattered |
| 2008 | SP266 | 10 | 9 | 10 | 21 | 41.2 | 0.13 | 19.5 | 5.8 | 3 | tno |
| **2010** | **RN64** | 10 | 9 | 10 | 21.4 | 40.93 | 0.06 | 19.9 | 5.7 | 82d | tno |
| **2010** | **RO64** | 10 | 9 | 10 | 21.2 | 47.06 | 0.13 | 17 | 5.3 | 3 | tno |

|  |  | | | | | | | | |  |
|---|---|---|---|---|---|---|---|---|---|---|
| **2010** | **TH**[e] | 10 | 10 | 2 | 20.2 | 18.65 | 0.33 | 26.7 | 9.2 | 5 | centaur |
| **2010** | **TJ** | 10 | 10 | 2 | 21.2 | 63.62 | 0.38 | 38.9 | 5.2 | 2 | scattered |
| 2007 | PS45 | 10 | 10 | 4 | 21.1 | 48.02 | 0.18 | 19.5 | 5.4 | 2 | tno |
| **2010** | **TR19** | 10 | 10 | 7 | 20.9 | 36.67 | 0.69 | 25.4 | 5.4 | 26d | scattered |
| 2008 | SO266 | 10 | 10 | 8 | 20.8 | 39.54 | 0.24 | 18.8 | 6 | 3 | tno |
| **2010** | **TY53** | 10 | 10 | 8 | 20 | 39.2 | 0.46 | 22.5 | 5.2 | 3 | scattered |
| **2010** | **TF182**[f] | 10 | 10 | 8 | 21.4 | 39.7 | 0.07 | 32.6 | 6.2 | 3d | ... |
| 2005 | SA278 | 10 | 10 | 9 | 21.2 | 94.04 | 0.65 | 16.3 | 6.2 | 9 | scattered |
| 2003 | UB313 | 10 | 10 | 9 | 18.5 | 68.01 | 0.44 | 43.8 | -1.2 | 20 | scattered |
| 2007 | RW10 | 10 | 10 | 14 | 20.8 | 30.32 | 0.3 | 36.1 | 6.6 | 8 | scattered |
| 2003 | UZ413 | 10 | 10 | 31 | 20 | 39.42 | 0.22 | 12 | 4.2 | 11 | tno |
| 2005 | RO43 | 10 | 10 | 31 | 21.1 | 29.03 | 0.52 | 35.4 | 7.3 | 7 | scattered |
| 2003 | UZ117 | 10 | 11 | 2 | 20.4 | 44.53 | 0.14 | 27.4 | 5.3 | 8 | tno |
| 2003 | VB12 | 10 | 11 | 2 | 20.7 | 545 | 0.86 | 11.9 | 1.6 | 10 | oort |
| **2010** | **VQ11** | 10 | 11 | 2 | 21.6 | 45.49 | 0.15 | 13.3 | 5.7 | 3 | tno |
| **2010** | **VR11** | 10 | 11 | 2 | 21.2 | 41.34 | 0.1 | 30.8 | 5.6 | 2 | tno |
| 2001 | PT13[e] | 10 | 11 | 3 | 19.4 | 10.67 | 0.2 | 20.3 | 9 | 10 | centaur |
| **2010** | **VW11** | 10 | 11 | 3 | 20.8 | 52.17 | 0.32 | 27.8 | 5.5 | 2 | scattered |
| **2010** | **VX11** | 10 | 11 | 3 | 20.8 | 42 | 0.49 | 22.4 | 6.3 | 57d | scattered |
| 2005 | RR43 | 10 | 11 | 5 | 19 | 43.5 | 0.14 | 28.5 | 4 | 8 | tno |
| **2010** | **VK201** | 10 | 11 | 5 | 21 | 43.4 | 0.11 | 28.8 | 4.5 | 2 | tno |
| 1998 | SM165 | 10 | 11 | 11 | 21.3 | 47.95 | 0.37 | 13.5 | 5.8 | 14 | tno |
| 1999 | TC36 | 10 | 11 | 11 | 19.6 | 39.67 | 0.23 | 8.4 | 4.9 | 9 | tno |
| 2003 | UR292 | 10 | 11 | 11 | 21.4 | 32.55 | 0.18 | 2.7 | 7.3 | 9 | tno |
| 2005 | RS43 | 10 | 11 | 11 | 21 | 48.04 | 0.2 | 10 | 5.3 | 11 | tno |
| 2002 | UX25 | 10 | 11 | 12 | 19.2 | 42.93 | 0.14 | 19.4 | 3.7 | 11 | tno |
| **2010** | **VL201** | 10 | 11 | 12 | 21.2 | 42.36 | 0.19 | 5.3 | 6.1 | 51d | tno |
| **2010** | **VZ98** | 10 | 11 | 12 | 20.3 | 134 | 0.74 | 4.5 | 5 | 2 | scattered |
| 2007 | TY430 | 10 | 11 | 13 | 20.8 | 39.5 | 0.27 | 11.3 | 6.9 | 4 | tno |
| 2004 | UX10[e] | 10 | 11 | 14 | 19.9 | 39.21 | 0.04 | 9.5 | 4.5 | 15 | tno |
| **2010** | **WG9** | 10 | 11 | 30 | 20.7 | 53.68 | 0.65 | 70.2 | 8.1 | 3 | scattered |

| | | | | | | | | | | |
|---|---|---|---|---|---|---|---|---|---|---|
| 2010 | **XS58**[f] | 10 | 12 | 3 | 21 | 19.2 | 0.06 | 16.2 | 8.9 | 2d | ... |
| 2010 | **XT58**[f] | 10 | 12 | 10 | 21.4 | 31.7 | 0.02 | 22.3 | 7 | 2d | ... |
| 2003 | AZ84 | 11 | 1 | 3 | 19.8 | 39.6 | 0.18 | 13.5 | 3.7 | 9 | tno |
| 2011 | **FW62**[e] | 11 | 3 | 25 | 21.2 | 39.83 | 0.17 | 26.8 | 5 | 62d | tno |
| 2011 | **FY9**[e] | 11 | 3 | 27 | 20.9 | 58.73 | 0.74 | 37.8 | 8.8 | 2 | scattered |
| 2011 | **FX62** | 11 | 3 | 28 | 20.3 | 47.9 | 0.58 | 18.3 | 7.4 | 5 | scattered |
| 2011 | **GM27** | 11 | 4 | 2 | 21.1 | 51.75 | 0.44 | 13.1 | 5.2 | 59d | scattered |
| 2011 | **GN27**[e] | 11 | 4 | 3 | 21 | 15.41 | 0.42 | 15.7 | 9.6 | 1d | centaur |
| 2004 | GV9 | 11 | 4 | 6 | 19.7 | 41.84 | 0.08 | 22 | 4 | 8 | tno |
| 2011 | **GP61**[e] | 11 | 4 | 8 | 20.1 | 11.9 | 0.14 | 15.2 | 10.7 | 5d | centaur |
| 2011 | **GY61** | 11 | 4 | 9 | 21.6 | 50.07 | 0.3 | 0.8 | 6.5 | 2 | scattered |
| 2011 | **HO60** | 11 | 4 | 26 | 21.4 | 91.5 | 0.61 | 22.3 | 6.2 | 63d | scattered |
| 2000 | GP183 | 11 | 4 | 29 | 21.5 | 39.67 | 0.07 | 4.9 | 6.6 | 4 | tno |
| 2011 | HP83[e] | 11 | 4 | 29 | 20.9 | 52.25 | 0.31 | 28.7 | 5.6 | 2 | scattered |
| 2011 | **HQ83**[f] | 11 | 4 | 30 | 21.9 | 38.6 | 0.03 | 8 | 6.1 | 10d | ... |
| 2011 | **JF31** | 11 | 5 | 3 | 21.3 | 41.32 | 0.14 | 27.8 | 5.6 | 2 | tno |
| 2011 | **JG31** | 11 | 5 | 4 | 20.3 | 40.3 | 0.14 | 3.5 | 5.6 | 22d | tno |
| 2012 | **HG84**[e] | 12 | 4 | 17 | 22.1 | 42.45 | 0.06 | 10.9 | 6.4 | 67d | tno |
| **LSQ** | **188**[e,f] | 12 | 4 | 17 | 21.3 | 55.1 | 0.02 | 12.24 | 4 | 1d | ... |

Notes: All orbital elements are from the MPC, unless noted otherwise; [a] LSQ discoveries are in bold font; [b] unit is number of oppositions, or days if followed by "d"; [c] based on the MPC listing, with "scattered disk/centaur" separated by aphelion greater or less than 30 AU, respectively; [d] redetected during the systematic survey, beginning 2010 August; [e] detected in a field not meeting the acceptance criteria of the survey; [f] orbital elements and H computed using routines from Bernstein & Khushalani (2000).

Figure Captions

1. Scale drawing showing the arrangement of CCDs in the 160-Megapixel QUEST camera. Non-functioning CCDs are represented by blank areas on the fingers. The arrows are vectors showing the amount and direction of the pointing offsets we use to increase the completeness of our sky coverage.

2. Transmission versus wavelength of the $Q_{st}$ interference filter. The oscillations in transmission are real features, resulting from the interference film required to manufacture the filter .

3. The discovery images of 2010 WG9 (top row, indicated by the white line in the first three images) and of a false source (bottom row). For each square panel , North is up and East to the left, and the scale is 88" x 88". See text for further explanations.

4. The single-pointing detection efficiency, $\varepsilon$, versus USNO R-band magnitude, $M_R$, determined from observations of known main-belt asteroids (open squares) and of known KBOs (filled squares). The solid line shows the efficiency function obtained by simultaneously fitting the main-belt observations for $M_R < 20.5$ and the KBO observations for $M_R > 20.5$. The parameters of the fit are listed in Table 1.

5. The dithered-pointing detection efficiency, $\kappa$, versus USNO R-band magnitude, $M_R$, determined from observations of known main-belt asteroids. This is the efficiency obtained by combining observations from 4 dithered pointings. The data have been fit with the same function used to fit $\varepsilon$ in Fig. 4 (see Eq. 1).

6. The areas of the sky searched by the La Silla – QUEST KBO survey (2010 August to 2012 May) that meet the acceptance criteria of the survey. Areas covered with a complete set of 4 dithers (cyan) have been searched with efficiency 84%. Other areas covered with partial dithers (blue) were searched with lower efficiency (in the printed version, cyan and blue appear as light and dark grey). The dashed and solid lines show the galactic and ecliptic plane, respectively.

7. The number of KBOs and Centaurs detected with the La Silla – QUEST KBO survey (solid line) and the fractional sky coverage versus ecliptic latitude (dashed line). Only fields meeting our acceptance criteria are included in the fractional sky coverage, and only objects detected in these accepted fields are included in the count.

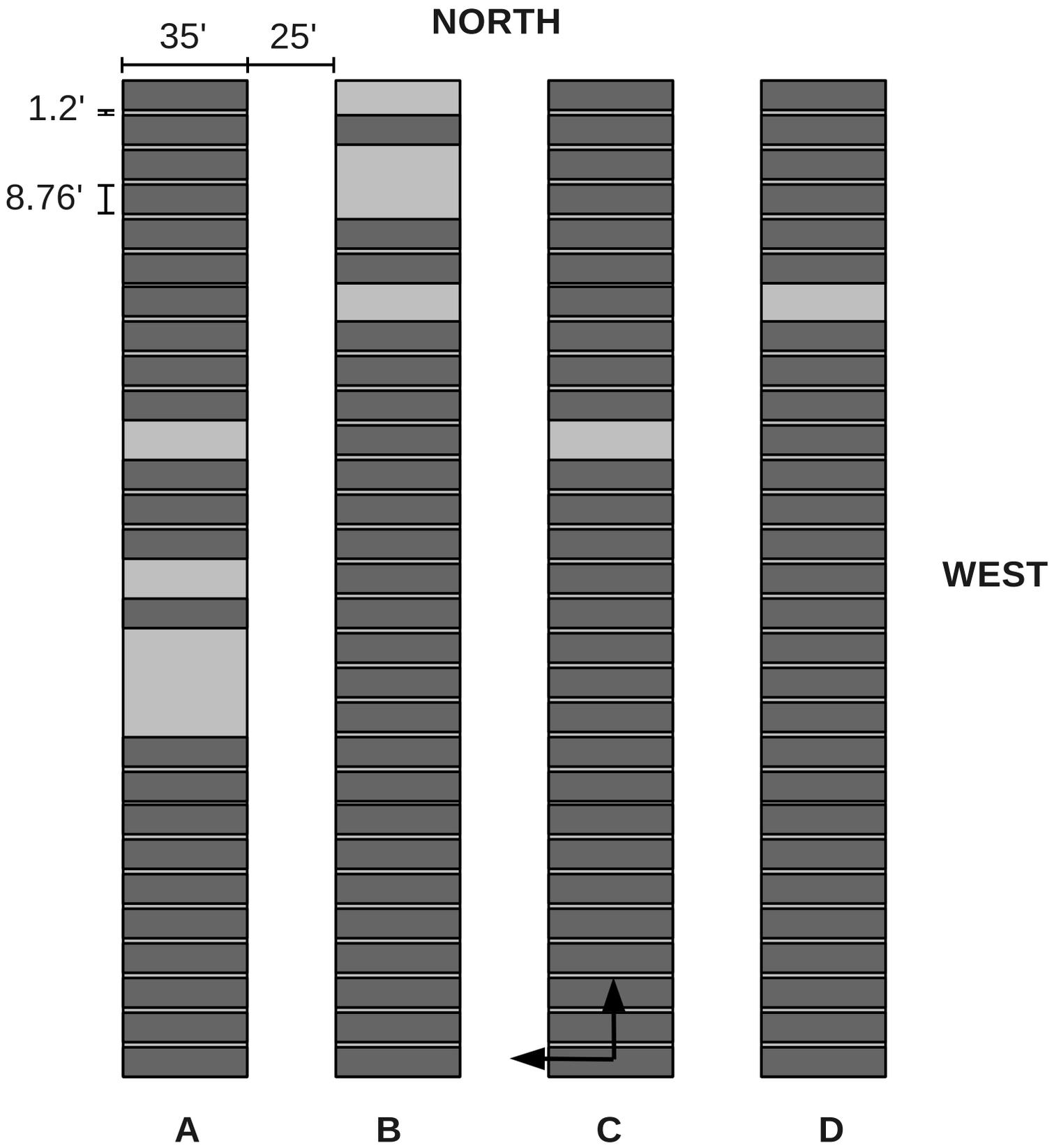

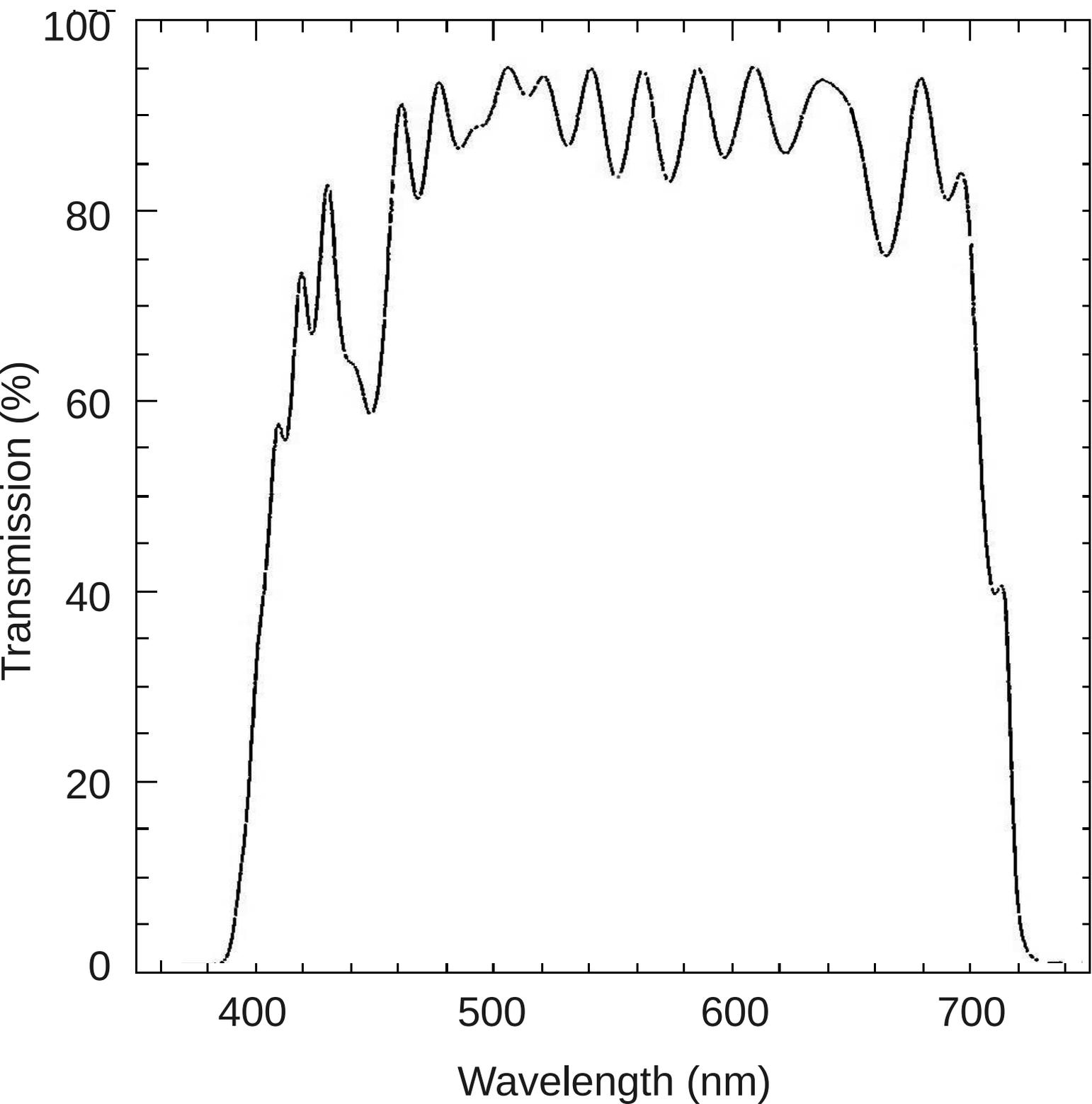

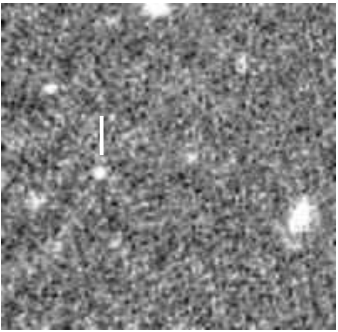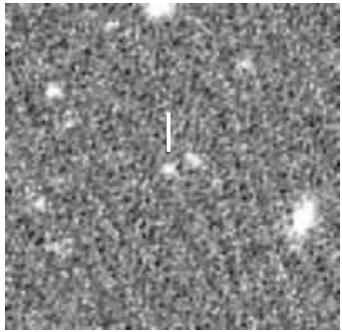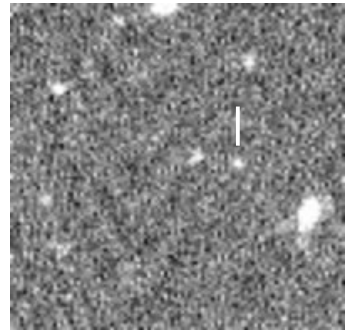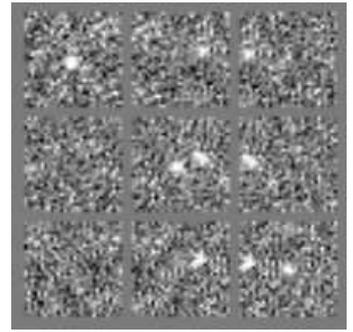
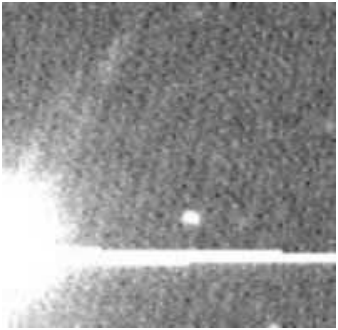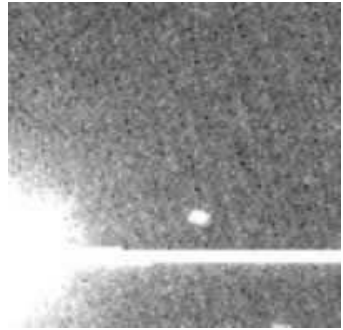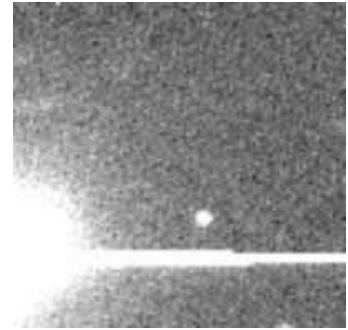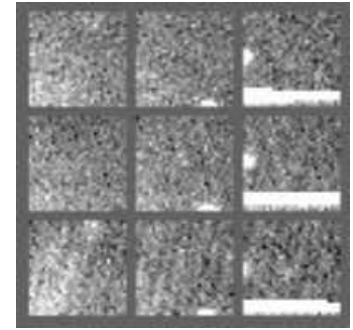

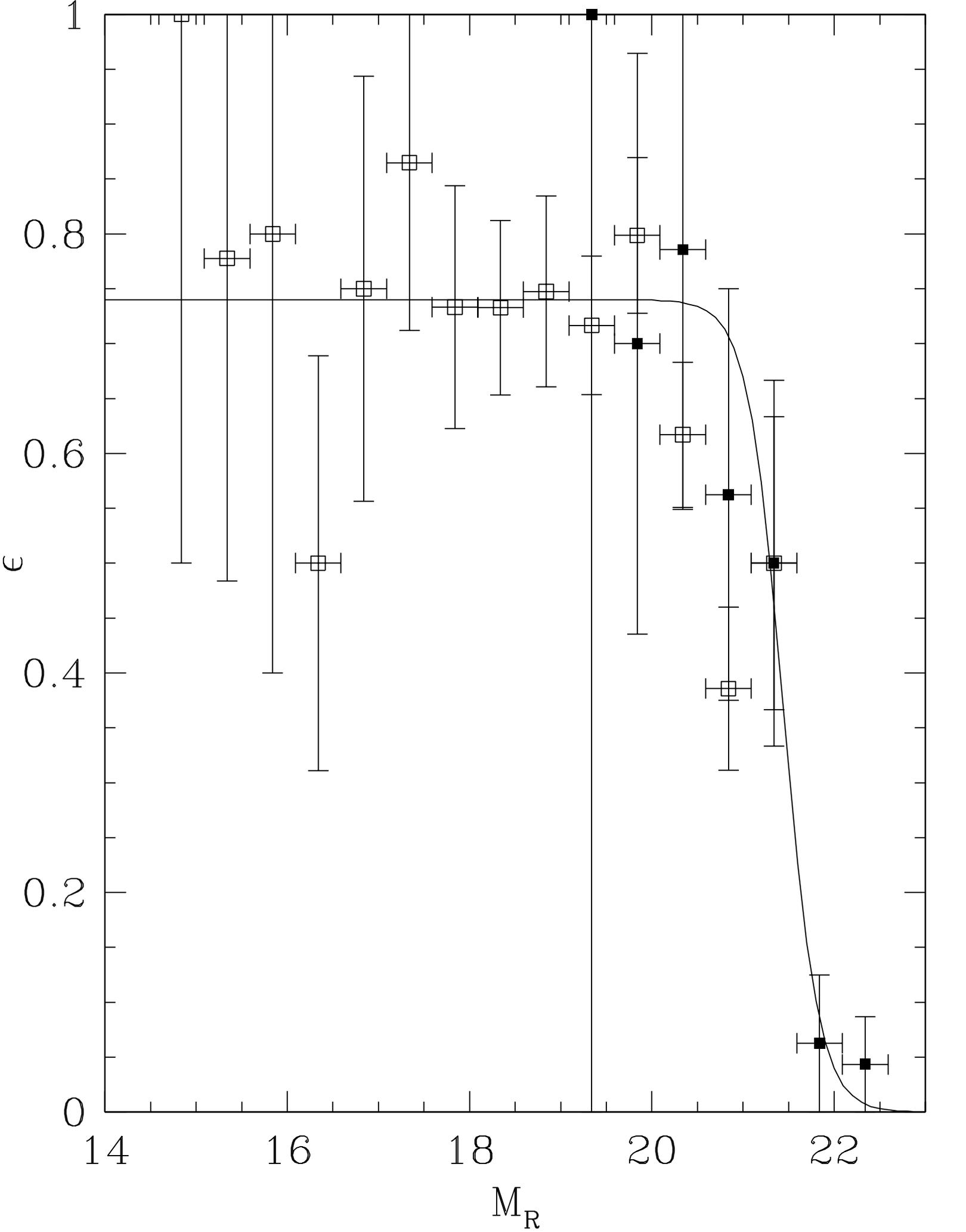

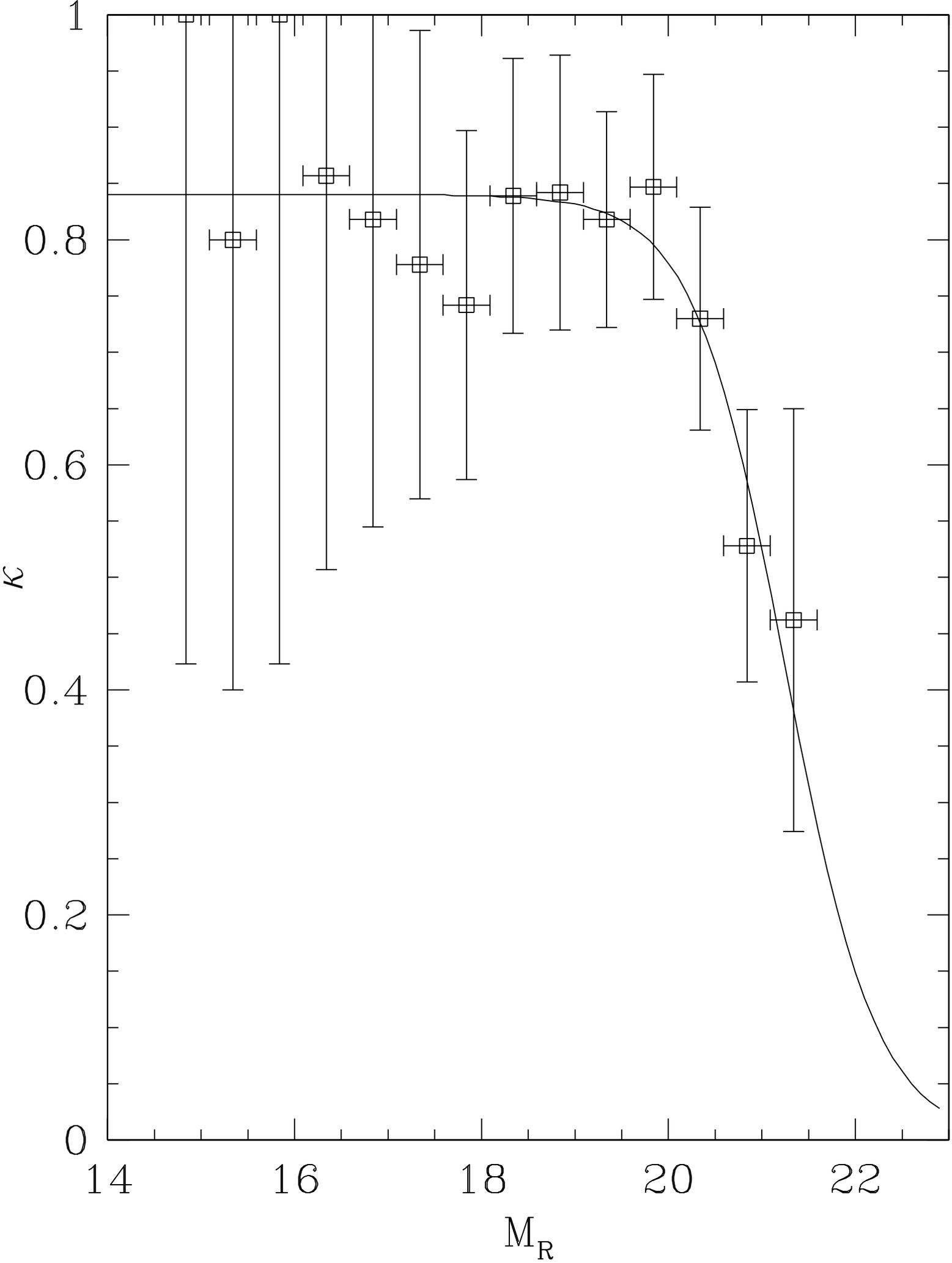

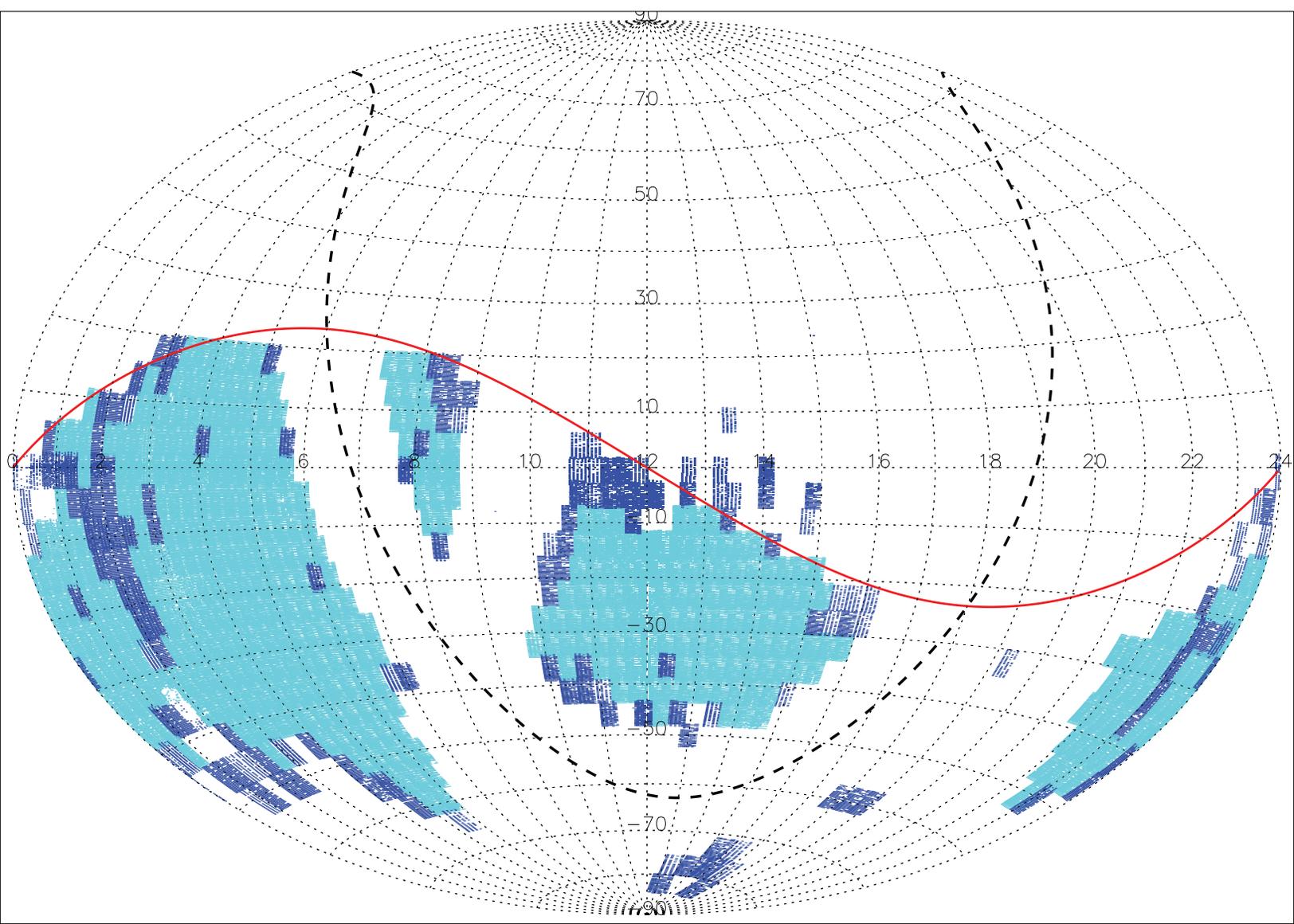

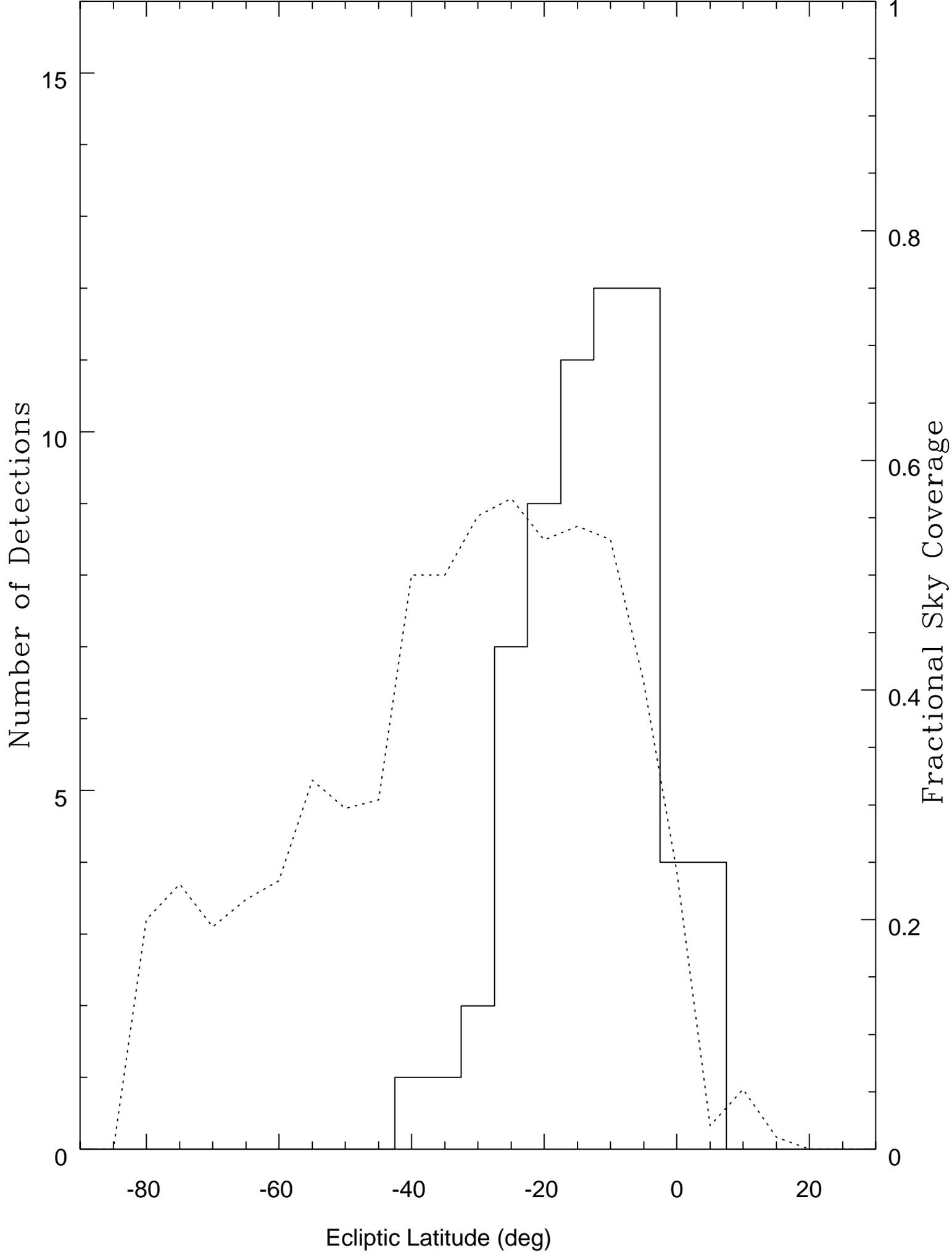